\documentclass[conference,a4paper]{IEEEtran}
\usepackage{cite}
\usepackage{amsmath,amssymb,amsfonts}
\usepackage{algorithmic}
\usepackage{graphicx}
\usepackage{textcomp}
\usepackage{threeparttable}
\usepackage{multirow}
\usepackage{colortbl}
\usepackage{tabularray}
\usepackage{booktabs}

\def\BibTeX{{\rm B\kern-.05em{\sc i\kern-.025em b}\kern-.08em
    T\kern-.1667em\lower.7ex\hbox{E}\kern-.125emX}}

\begin{document}

\title{Low-Power PLL-Based Clock Stabilization for Flexible IGZO AMS Systems}

\author{\IEEEauthorblockN{Paula Carolina Lozano Duarte\IEEEauthorrefmark{1},
Georgios Zervakis\IEEEauthorrefmark{2}, \IEEEmembership{Member, IEEE} and
Mehdi Tahoori\IEEEauthorrefmark{1}, \IEEEmembership{Fellow, IEEE}}
\IEEEauthorblockA{\IEEEauthorrefmark{1}Dept. of Computer Science, Karlsruhe Institute of Technology, Karlsruhe 76131, Germany}
\IEEEauthorblockA{\IEEEauthorrefmark{2}School of Electrical and Computer Engineering, National Technical University of Athens, Athens 15772, Greece}
\thanks{This work has been supported by the European Research Council (ERC) (Grant No. 101052764)}
\thanks{Corresponding author: Paula Carolina Lozano Duarte (e-mail: paula.duarte@kit.edu).}}

\maketitle

\begin{abstract}
Flexible electronics (FE) platforms rely on analog and mixed-signal (AMS) circuits-biosensors, readout front-ends, and analog to digital converters-that dominate both functionality and energy consumption, making on-chip clock generation an essential yet power-critical function.
Existing oscillator-based solutions suffer from unbounded process, voltage, and temperature (PVT) drift that degrades signal integrity, while alternative clock sources can consume up to 90\,\% of the total system power budget-rendering them inapplicable to FE platforms and elevating clock generation to a primary power and energy-efficiency design constraint.
This paper presents the first phase-locked loop (PLL) architecture designed for n-type-only amorphous indium-gallium-zinc oxide (a-IGZO) thin-film transistor (TFT) technology, addressing FE-specific constraints such as the absence of p-type devices, limited carrier mobility, and strong PVT variability.
Rather than targeting high-precision frequency synthesis, the proposed design operates as a low-bandwidth temporal stabilizer: a free-running ring-oscillator-based voltage-controlled oscillator (VCO) is softly regulated by a minimal charge-pump feedback loop to bound long-term frequency drift without requiring a continuous high-quality external reference.
The proposed PLL supports frequencies from 1\,kHz to 300\,kHz while occupying 0.0115--0.0233\,mm$^{2}$ and consuming 0.115--0.153\,mW. 
Compared with prior oscillator-based FE clocking solutions, our architecture reduces power by more than 400$\times$ while achieving footprint reductions exceeding 1500$\times$ compared to flexible VCOs, and more than 390$\times$ with respect to ring-oscillator solutions.
Validated across four representative published IGZO AMS systems, the proposed PLL achieves an rms period jitter of 2.24\,ns and a long-term frequency accuracy within 1000\,ppm, providing reference-anchored clock stability in FE platforms.
\end{abstract}

\begin{IEEEkeywords}
Phase-locked loop, Flexible Electronics, IGZO TFT, low-power clock generation, AMS systems
\end{IEEEkeywords}

\section{Introduction}\label{sec:intro}

Flexible electronics (FE) enable wearable and large-area sensing systems where ultra-low power operation and monolithic integration are key design objectives~\cite{tahoori2025computing}. 
Recent flexible and IGZO-based platforms demonstrate sensing and processing pipelines operating at kHz frequencies with milliwatt-to-sub-milliwatt power budgets~\cite{Gao:FlexibleWearableSensing}.
In these systems, analog and mixed-signal (AMS) blocks — such as sensors, readout circuits, and analog-to-digital converters (ADCs) — dominate both functionality and energy consumption, making efficient clock generation a critical yet often overlooked system component~\cite{Afentaki:LES2024}.

Clock generation is almost universally treated as an afterthought: most published IGZO AMS systems report power budgets that do not account for the clock generator, implicitly assuming an external source or a free-running ring oscillator (RO).
This assumption is overly optimistic with respect to total system hardware overhead and integration complexity.
For example, when a state-of-the-art IoT CMOS Phase  Locked Loop (PLL)~\cite{Liu:TCSI2017:IoTPLL} is integrated into representative IGZO systems (Section~\ref{sec:results}), the clock alone consumes up to 90\,\% of the total system power---far above the clock-network power share exceeding 25\,\% commonly reported in conventional silicon systems~\cite{Friedman:ClockDistribution}.
The clock generator is therefore a primary determinant of system-level energy efficiency rather than a peripheral block.

Beyond power, free-running ROs suffer from unbounded PVT drift: in the kHz operating regime of FE AMS systems, moderate threshold-voltage shifts can translate into frequency errors exceeding 30,\%, directly degrading ADC sampling accuracy and forcing conservative timing guard-bands~\cite{Papadopoulos:C-2CSAR}.
Existing FE clocking solutions rely almost exclusively on open-loop oscillators, which inherently lack bounded frequency accuracy.
The key missing building block is not a high-precision synthesizer but a low-power mechanism to bound temporal uncertainty over time.
Designing such a clock generator in FE technologies introduces additional constraints compared to CMOS circuits, including the absence of p-type devices, limited carrier mobility, large threshold-voltage variability, and operation at low frequencies (kHz range).

We propose a minimal charge-pump PLL tailored to these constraints, where a low-bandwidth feedback loop softly regulates a free-running RO-based Voltage-Controlled Oscillator (VCO) to replace open-loop drift with reference-anchored stability.
The proposed architecture operates at 3\,V supply and supports frequencies from 1\,kHz to 300\,kHz by adjusting only the VCO load capacitance, without transistor resizing or topology changes. 
Across this range, the PLL occupies 0.0115--0.0233\,mm$^{2}$ and consumes 0.115--0.153\,mW while providing bounded frequency accuracy of 1000\,ppm and rms jitter of 2.24\,ns. 
Compared with prior oscillator-based FE clocking solutions operating at 5--25\,V, the proposed architecture reduces power by more than 400$\times$ and footprint by more than 390$\times$.
To demonstrate practical integration, the PLL topology is evaluated across four representative IGZO AMS systems spanning 1\,kHz to 300\,kHz, showing that closed-loop clock stabilization can be incorporated with modest silicon and energy overhead.
All results are verified through post-layout simulation using PragmatIC's FlexIC PDK, confirming that parasitic effects introduced by the physical layout do not degrade the reported performance metrics.
\textbf{The key contributions of this work are the following:}

\begin{itemize}

    \item The first PLL implemented in n-type-only IGZO TFT technology, achieving 1--300\,kHz operation at 3\,V with 0.115--0.153\,mW power and 0.0115--0.0233\,mm$^{2}$ area.
    \item Frequency-scalable architecture via VCO load capacitance tuning without transistor resizing.
    \item Replacement of open-loop oscillator drift with bounded 1000\,ppm clock stability in FE platforms.
    \item System-level validation across four published IGZO AMS systems demonstrating practical integration of closed-loop clock stabilization in FE platforms.

\end{itemize}

\section{Background}\label{sec:background}

\subsection{Flexible Electronics and IGZO Technology}\label{sec:FE}

Flexible electronics (FE) encompasses devices built on deformable substrates that maintain operational integrity while being bent, stretched, or twisted.
This mechanical adaptability facilitates deployment on non-planar surfaces, including biological tissues, textiles, and soft-robotic components, positioning FE as a particularly suitable technology for wearable health monitoring systems, conformable sensing devices, and interactive human-machine interfaces~\cite{Gao:FlexibleWearableSensing, afentaki2025islped}.
In contrast to conventional rigid silicon-based technologies, FE devices must operate in environments demanding mechanical compliance, cost-effectiveness, and resilience to physical stress.

Indium-gallium-zinc oxide (IGZO) constitutes a key enabling material in FE, serving as a wide-bandgap semiconductor in thin-film transistor (TFT) technologies.
IGZO-TFTs exhibit relatively high electron mobility, good uniformity, and compatibility with low-temperature deposition processes, making them well-suited for large-area, low-power applications~\cite{Zhu:IGZO2021, Pan:IGZOTFT2024}.
The absence of complementary (p-type) devices represents a technological limitation that frequently results in increased static power consumption and necessitates alternative design strategies, such as dynamic logic or pseudo-CMOS configurations, for implementing logic and analog circuits.

Despite these constraints, IGZO-based FE circuits have demonstrated the feasibility of integrating essential analog and mixed-signal building blocks, such as operational amplifiers~\cite{Zysset:opamptft, Meng:opampigzo}, signal conditioning stages~\cite{Garripoli:analogfrontendigzo, lozano:AKAN}, and various ADC architectures~\cite{Papadopoulos:C-2CSAR, Lozano:aspdac25:BinCoDesign}.
More recently, IGZO integration has been extended to digital processing, with flexible RISC-V processors~\cite{FlexRISCV:MICRO25, FlexCoProc:DATE26, Ozer2024BendableRISCV} and mixed-signal machine-learning inference engines~\cite{FlexClassifier:ICCAD25, afentaki:date26:mixed_signal} demonstrating system-level AMS functionality---all operating at 3\,V and requiring well-defined clocks in the 1\,kHz to 300\,kHz range.
FE does not aim to replace silicon-based platforms, but rather to complement them in domains where low cost, flexibility, and conformality are application-critical requirements~\cite{Luo:smartpack, lee:flexiblepatch}.

\subsection{Clock Generation in FE}
\label{sec:clocks_fe}

Clock generation is a foundational requirement for any digital or mixed-signal system, yet it has received comparatively little attention in the FE literature.
The dominant approach in reported IGZO systems is to rely on an off-chip clock source or, where full integration is required, to employ a free-running ring oscillator (RO) built from a chain of inverting stages.

Ring oscillators are naturally compatible with n-type-only IGZO technologies: a pseudo-CMOS inverter chain can be constructed using only n-type TFTs with diode-connected or bootstrapped loads, and the oscillation frequency is set by the propagation delay of each stage~\cite{Zhu:IGZO2021, Pan:IGZOTFT2024}.
Their main appeal is extreme simplicity---no passive components or references are needed---and the target frequency can be adjusted by varying the number of stages or the load capacitance.
However, because frequency is determined by transistor threshold voltage and carrier mobility, both of which are highly sensitive to process spread, operating temperature, and supply voltage, ROs exhibit large and unbounded PVT-induced drift~\cite{tahoori2025computing}.
In the kHz-range operating regime of FE AMS systems, even a modest threshold-voltage shift of a few hundred millivolts can translate into frequency errors exceeding tens of percent, directly degrading ADC sampling accuracy and increasing the energy cost of conservative timing guard-bands~\cite{lozano2026faulttolerantdesignigzobased}.

RC-based relaxation oscillators offer an alternative in which frequency is set by a resistor-capacitor time constant, partially decoupling oscillation period from transistor parameters.
Although this improves temperature stability relative to ROs, absolute frequency accuracy remains limited by the tolerances of on-chip passive components and by the comparator threshold, which is itself device-dependent in IGZO.
Digital frequency calibration---in which a counter-based scheme periodically adjusts a digitally controlled oscillator---has been explored to reduce PVT sensitivity in low-power mixed-signal contexts~\cite{Papadopoulos:C-2CSAR}, but adapting such schemes to IGZO requires a calibration controller and reference counter that conflict with the area and power budgets of FE platforms, and it corrects only quasi-static drift rather than cycle-to-cycle jitter.

Neither ROs nor RC oscillators provide any mechanism to continuously bound accumulated phase error over time, leaving long-term frequency stability---critical for synchronization across multi-block AMS pipelines---fundamentally unaddressed.
To the best of our knowledge, \textit{no PLL has yet been proposed in any metal-oxide or organic TFT technology, making the present work the first attempt to bring closed-loop clock stabilization to FE.}

\subsection{Phase-Locked Loops in CMOS}\label{sec:plls_cmos}

The charge-pump PLL is a well-established CMOS building block comprising a Phase-Frequency Detector (PFD), Charge Pump (CP), Low Pass Filter (LPF), VCO, and frequency divider~\cite{Razavi:DesignAnalogCMOS}.
Representative IoT designs achieve GHz-range synthesis below 2\,mW in advanced CMOS~\cite{Liu:TCSI2017:IoTPLL, Tavakoli:JLPEA2023:IntNPLL}.
Porting these architectures to IGZO faces four fundamental obstacles: (i)~the absence of p-type devices eliminates complementary PFD and charge-pump topologies, requiring n-type-only alternatives with additional static current;
(ii)~IGZO electron mobility (5--20\,cm$^{2}/(V\!\cdot\!s))$ is orders of magnitude below CMOS, severely limiting achievable loop bandwidth; 
(iii)~threshold-voltage variability across flexible substrates far exceeds the CMOS PVT window, requiring the VCO tuning range to be sized conservatively; and
(iv)~kHz-range loop poles require loop-filter capacitors in the picofarad-to-nanofarad range---orders of magnitude larger than their CMOS counterparts---motivating careful topology selection.

\subsection{Motivation}
\label{sec:background:motivation}

Most IGZO AMS publications report power consumption that excludes the clock generator; the ``complete system'' power cited in comparisons is therefore systematically understated.
For example, when a state-of-the-art IoT CMOS PLL~\cite{Liu:TCSI2017:IoTPLL} is
integrated into representative IGZO systems, the clock alone accounts for 30--90\,\% of total system power. 
For context, clock networks in silicon designs account for $\approx$25\,\% of chip power~\cite{Friedman:ClockDistribution}; in IGZO, the ratio is far higher because the host systems themselves are highly power-frugal.
This argument extends to any unoptimized clock source: a free-running RO sized for the same frequency would consume comparable power while providing no stability guarantee.
A dedicated low-power IGZO PLL is therefore a power- and energy-efficiency requirement.

\section{Proposed Flexible PLL}\label{sec:metho}

\subsection{System Context and Target Specifications}
\label{sec:metho:context}

\begin{figure}
    \centering
    \includegraphics[width=\linewidth]{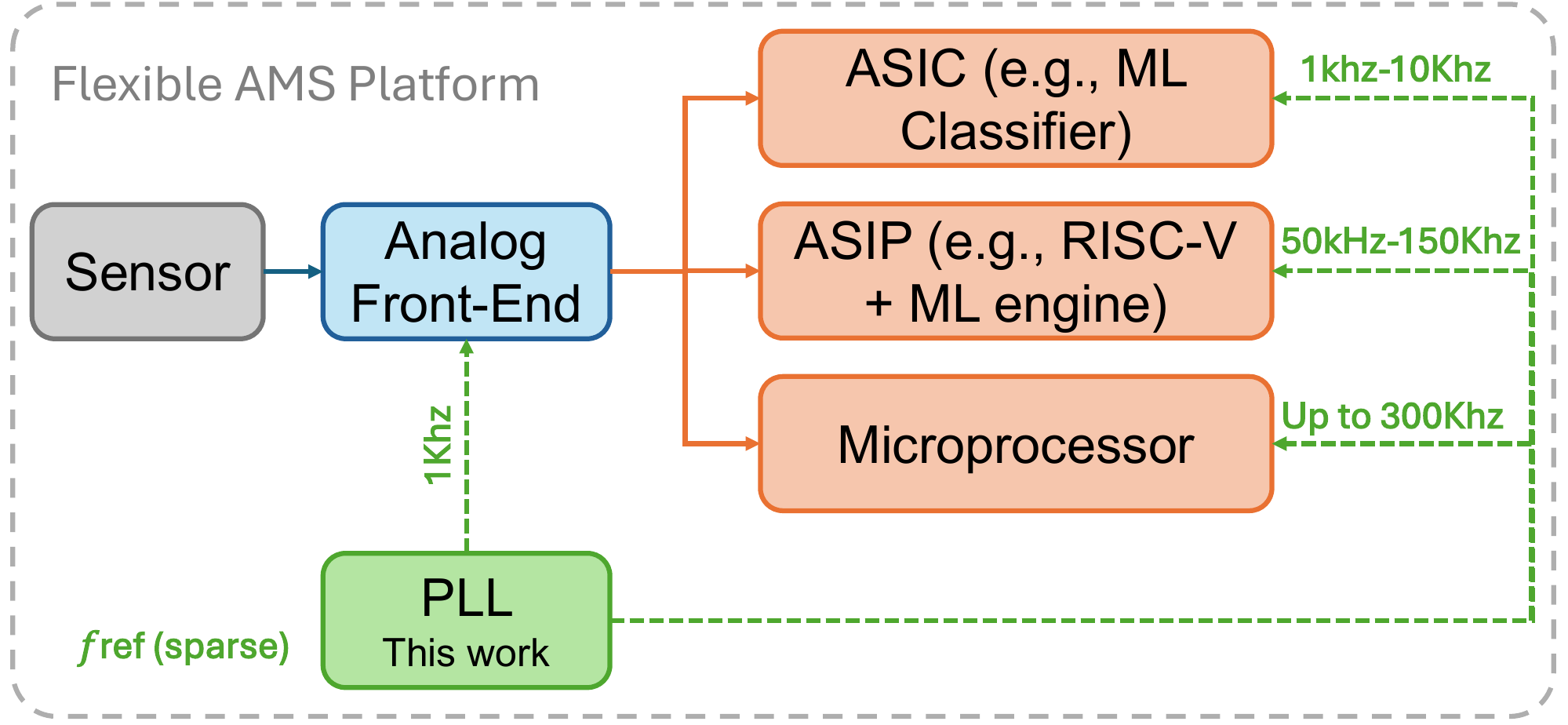}
    \caption{Conceptual integration of the proposed PLL within a flexible AMS platform. Sensor data is processed by the analog front-end and forwarded to different digital processing blocks (ASIC, ASIP, and microprocessor), each operating at distinct frequency ranges. The PLL provides clock signals across these domains, while $f_\mathrm{ref}$ is only intermittently required. Only the VCO load capacitor is adjusted to set the target frequency.}
    \label{fig:conceptual} 
\end{figure}

The proposed PLL is conceived as an \emph{integrated clocking building block} for flexible IGZO AMS platforms.
To identify typical system requirements before designing, we review representative IGZO systems from the recent literature (Table~\ref{tab:targets}), all operating at 3\,V—the most common supply in IGZO literature, which avoids level-shifting overhead—and spanning nearly three decades of clock frequency (1\,kHz–300\,kHz).
Together, they highlight the range of area, power, and frequency constraints encountered in wearable electronics, covering the two dominant application classes: digital processing~\cite{FlexRISCV:MICRO25, FlexCoProc:DATE26} and mixed-signal inference~\cite{FlexClassifier:ICCAD25, Alkhalil:BioCAS:2022:FlexibleSAR}.
Fig.~\ref{fig:conceptual} illustrates how a single stabilized clock can be shared across multiple target systems, emphasizing the reusability of the PLL.

\begin{table}[h]
  \centering
  \caption{Target IGZO AMS systems ($V_\mathrm{DD}=3$\,V).}
  \label{tab:targets} 
  \renewcommand{\arraystretch}{1.25}
  \begin{tabular}{lccc}
    \toprule
    \textbf{Ref.} &
    $f_\mathrm{clk}$ &
    \textbf{Area (mm$^2$)} &
    \textbf{Power (mW)} \\
    \midrule
    \cite{FlexRISCV:MICRO25}                & 300\,kHz        & 0.59--1.02 & 0.71--0.91 \\
    \cite{FlexCoProc:DATE26}                & 150\,kHz        & 2.41--2.44 & 1.49--1.53 \\
    \cite{FlexClassifier:ICCAD25}           & 10\,kHz         & 0.20--27.6 & 0.14--20.3 \\
    \cite{Alkhalil:BioCAS:2022:FlexibleSAR} & $\approx$1\,kHz & 9.4        & 2.93       \\
    \bottomrule
  \end{tabular} 
\end{table}

Two guiding principles govern the design.  
\emph{Frequency co-design}: the VCO is tuned to the required application frequency by adjusting only the load capacitance $C$, while the PFD, charge pump, and all transistor sizes remain unchanged, enabling easy frequency adjustment and reusability across different targets.  
\emph{Overhead minimality}: the PLL must contribute negligible area and power relative to the host system, as any significant overhead would compromise the feasibility of flexible, conformal devices—for instance, by increasing patch size, shortening battery lifetime, or preventing self-powered operation.

\subsection{PLL Architecture}
\label{sec:metho:arch}

\begin{figure}
    \centering
    \includegraphics[width=\linewidth]{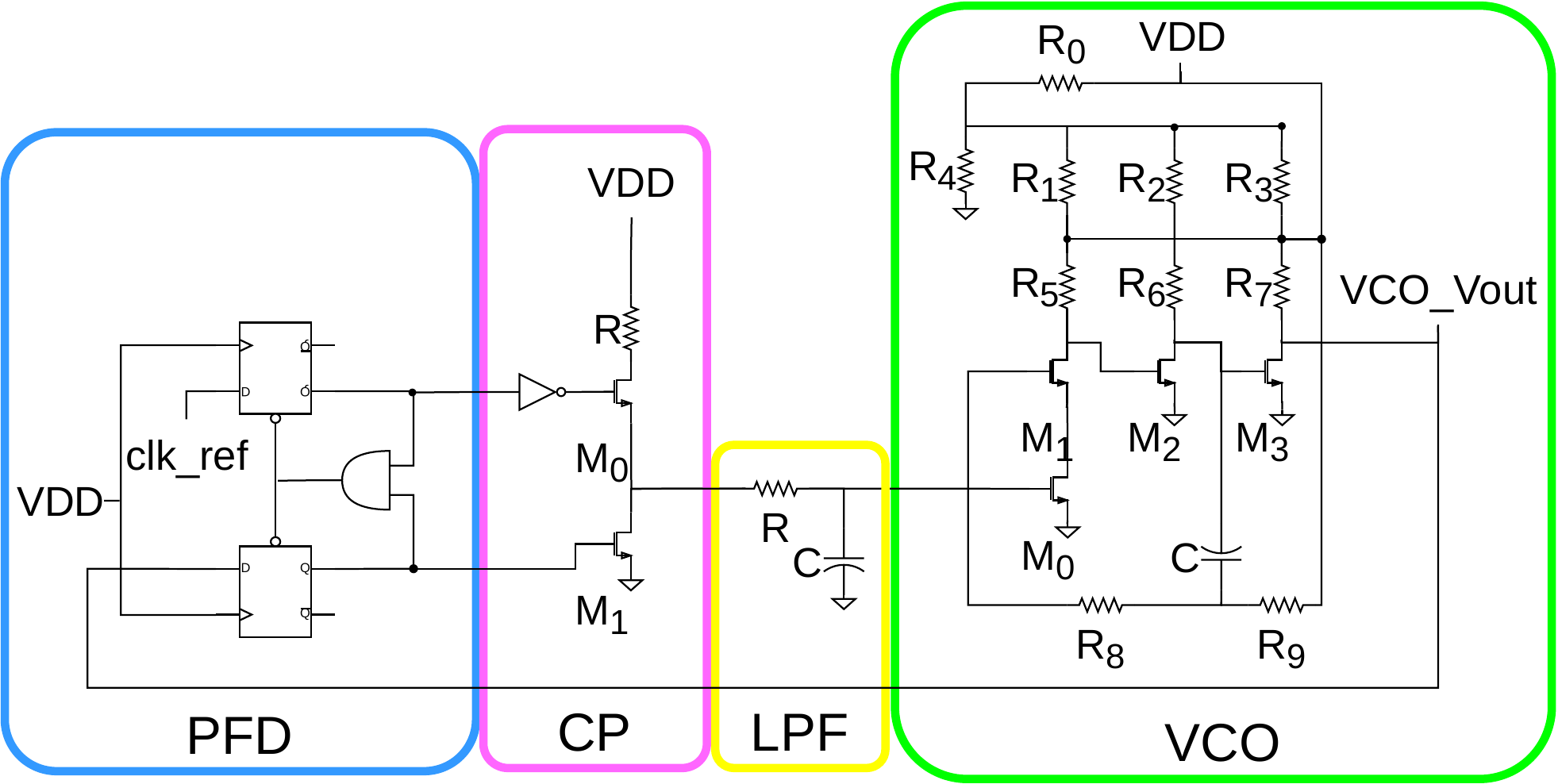}
    \caption{Proposed PLL schematic. Left: Phase Frequency Detector (PFD) built from two pseudo-CMOS dynamic D flip-flops (\texttt{DQQ}) and \texttt{clk\_ref} input. Center: charge pump (CP) and first-order low-pass filter (LPF). Right: three-stage ring-oscillator-based Voltage Control Oscillator (VCO) with per-stage resistive loads (R1--R7) and $V_\mathrm{ctrl}$-gated bias network (R0, R4, R8, R9, M0--M3).}
    \label{fig:PLL} \vspace{-2ex}
\end{figure}

Given the kHz operating range of the target systems and the constraints of IGZO TFT technology, a compact charge-pump PLL architecture is adopted.
The proposed PLL follows the classical topology (Fig.~\ref{fig:PLL}), comprising a phase-frequency detector (PFD), charge pump (CP), low-pass filter (LPF), and ring-oscillator VCO.
While this topology is well-established in CMOS~\cite{Razavi:DesignAnalogCMOS}, no prior work has demonstrated a closed-loop PLL in n-type-only IGZO TFT technology; the novelty lies in the full-system adaptation of each sub-block to the n-type-only, low-mobility, high-PVT-variability constraints of IGZO, and in the co-design and sizing of all blocks as an integrated clocking solution for flexible AMS platforms.
A frequency divider is intentionally omitted: the VCO output is locked directly to the reference, since kHz-range clocks can be generated directly by the ring oscillator, eliminating divider overhead.
All sub-blocks are implemented exclusively with n-type IGZO TFTs using pseudo-CMOS circuit techniques.

\subsubsection{Ring-Oscillator VCO}
Ring-oscillator-based VCOs are a well-known topology~\cite{Razavi:DesignAnalogCMOS}, and current-starved variants have been reported in unipolar TFT technologies~\cite{VCO_igzo_tejaswini}.
The novelty here lies in the resistive pseudo-CMOS stage topology and the single-capacitor frequency-scaling scheme, which together enable n-type-only operation and one-parameter retargeting across three decades of frequency without any transistor resizing.
The VCO is a three-stage ring oscillator in which each stage employs a resistive pseudo-CMOS topology: a resistive pull-up (R1–R7, 172.5\,k$\Omega$ each) and an n-type TFT pull-down driven by the preceding stage.
A bias network (R0, R4, R8, R9, M0–M3) controlled by $V_\mathrm{ctrl}$ regulates the stage current: increasing $V_\mathrm{ctrl}$ raises the bias current, reduces propagation delay, and increases oscillation frequency.
When retargeting, the load capacitor $C$ is the sole tuning parameter, enabling the topology to cover 1\,kHz to 300\,kHz.

\subsubsection{Phase-Frequency Detector}
The classical two-DFF charge-pump PFD is a standard CMOS building block~\cite{Razavi:DesignAnalogCMOS}.
Its implementation here is novel in that it is realized exclusively with pseudo-CMOS dynamic D flip-flops, avoiding any p-type device, and is sized to tolerate the VCO frequency spreads exceeding 30\,\% that are characteristic of IGZO PVT variability.
The PFD compares rising edges of $\mathrm{CLK}_\mathrm{ref}$ and the VCO output, generating UP/DOWN pulses accordingly.
A tri-state PFD is preferred over an XOR detector because it provides both phase and frequency detection, allowing lock acquisition from the large initial offsets caused by IGZO PVT spread.

\subsubsection{Charge Pump and Low Pass Filter}
A passive first-order CP + LPF combination is the standard choice in low-complexity PLLs~\cite{Razavi:DesignAnalogCMOS}.
The contribution here is the zero-quiescent-bias charge-pump implementation in n-type-only IGZO: eliminating any static current path is essential in a technology where transistor leakage and sizing options are limited, and directly enables the sub-0.2\,mW total system power reported in this work.
The charge pump converts UP/DOWN pulse widths into a net current that charges or discharges the passive first-order RC low-pass filter, producing $V_\mathrm{ctrl}$ while attenuating phase-detector ripple.
The pump current $I_\mathrm{CP}$ is sized to balance locking speed against $V_\mathrm{ctrl}$ ripple, which would otherwise modulate the VCO frequency and increase output jitter.

\subsubsection{Loop Dynamics and Bandwidth}
Narrowing the loop bandwidth to a fraction of the reference frequency is a standard PLL design principle~\cite{Razavi:DesignAnalogCMOS}.
What is specific to this work is the explicit co-design of the bandwidth constraint with IGZO's dominant impairments: the narrow bandwidth simultaneously suppresses low-frequency TFT flicker noise, reduces charge-pump switching activity, and provides slow-drift tracking—three requirements that arise directly from the low transconductance and strong PVT variability of IGZO TFTs. 
The loop bandwidth $f_\mathrm{BW}$ is kept at approximately one-tenth of $f_\mathrm{ref}$, maintaining adequate phase margin while limiting noise coupling into $V_\mathrm{ctrl}$.
For lower-frequency variants, the loop-filter RC values are re-dimensioned to preserve this ratio while all transistor dimensions remain unchanged.

\subsubsection{Reference Clock Requirements}
The reference input $\mathrm{CLK}_\mathrm{ref}$ can be provided by any stable clock source within the 1--300\,kHz range, such as a low-power crystal oscillator~\cite{lowpowerCrystal1} or a temperature-compensated oscillator module~\cite{TempCompensatedCrystal}.
Since the proposed PLL operates at kHz frequencies, standard 32.768\,kHz watch crystals~\cite{32kCrystal2} are directly compatible with the lower-frequency variants; a simple prescaler is sufficient to derive 1\,kHz or 10\,kHz references.
An alternative approach---dividing down a MHz-range crystal to the target frequency---would require an on-chip frequency divider, which in n-type-only IGZO adds pseudo-CMOS logic stages that increase area and static power.
Furthermore, since the proposed PLL operates without a divider, the VCO is locked directly to the reference, simplifying loop dynamics and reducing component count.
For the 300\,kHz variant, a dedicated 300\,kHz crystal or a divided-down 32.768\,kHz reference with phase interpolation provides sufficient accuracy for wearable AMS applications.
In all cases, the reference is only required intermittently, as the narrow loop bandwidth allows the PLL to maintain lock during brief reference interruptions---an important property for duty-cycled wearable systems.
\section{Experimental Setup and Results}\label{sec:results}

\subsection{Simulation Setup}
\label{sec:results:setup}

\begin{figure}[t]
    \centering
    \includegraphics[width=\linewidth]{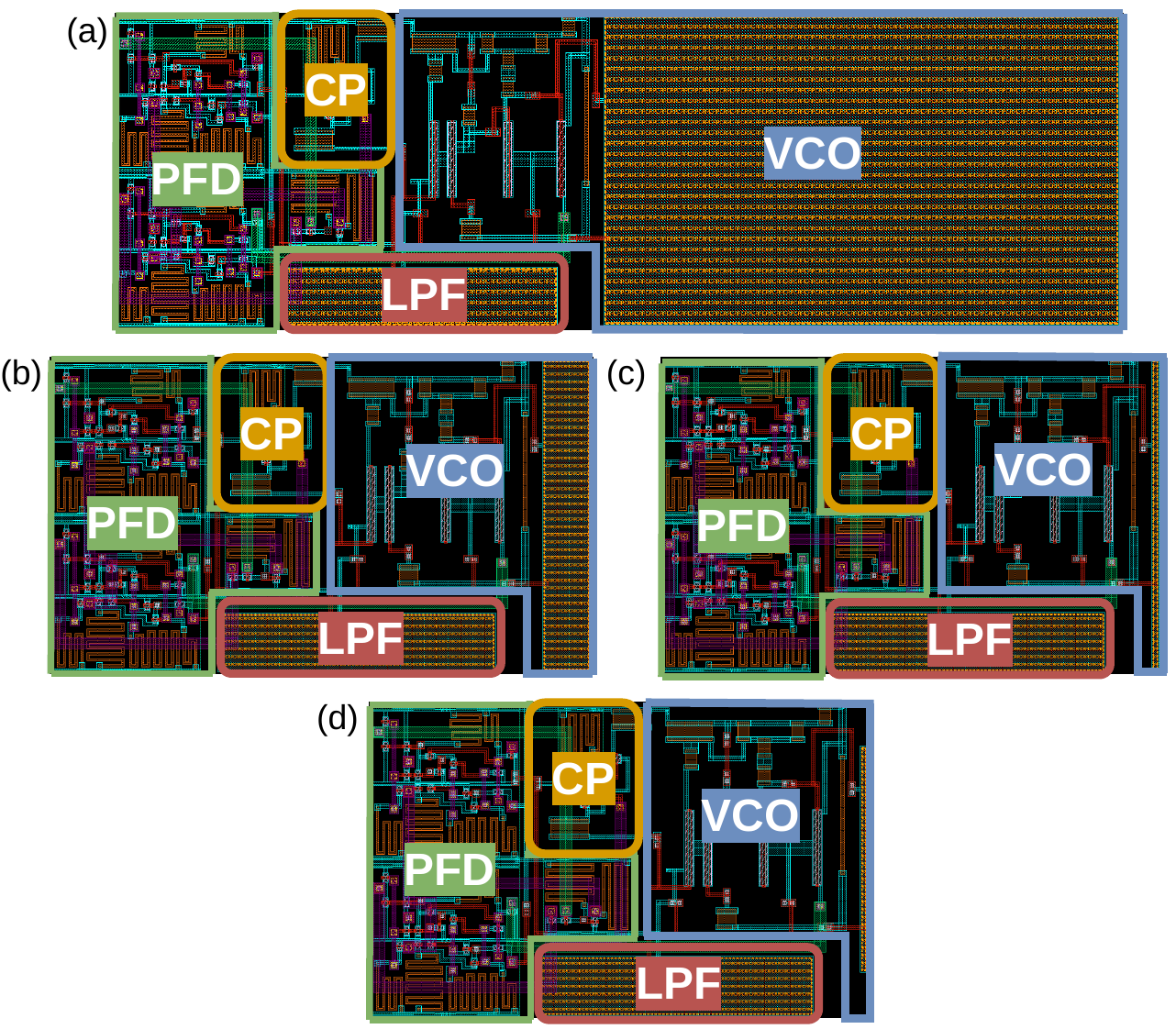}
    \caption{Physical layout of the PLL for each target frequency: (a)~1\,kHz (0.0233\,mm$^2$), (b)~10\,kHz (0.0125\,mm$^2$), (c)~150\,kHz (0.0115\,mm$^2$), and (d)~300\,kHz (0.0115\,mm$^2$). Area scales with the VCO load capacitor $C$; the 150\,kHz and 300\,kHz variants share the same footprint as only the loop-filter values differ.}
    \label{fig:layouts} 
\end{figure}

All simulations use Cadence Spectre with device models from PragmatIC's 3rd-generation FlexIC PDK~\cite{EuropracticeFlexibleElectronics}.
Physical layouts were completed for all four frequency variants, and all results reported in this section are obtained from \emph{post-layout} simulations that include extracted parasitic resistances and capacitances, ensuring that layout-dependent effects such as routing parasitics and coupling capacitances are fully accounted for. Their implementation is presented in Fig.~\ref{fig:layouts}.
Transient analysis over a 20\,ms window characterizes locking behavior, jitter, and power; all metrics are extracted over the locked interval (5--20\,ms) to exclude the acquisition transient.
Per-block power is obtained by inserting independent current probes on each sub-block supply rail.
PVT robustness is evaluated across three supply voltages (2.7, 3.0, 3.3\,V), three process corners (TT, FF, SS), and three temperatures (0\,$^\circ$C, 27\,$^\circ$C, 45\,$^\circ$C) targeting the wearable operating envelope, yielding 27 PVT combinations in total.
The temperature range was selected to cover standard room temperature (27\,$^\circ$C) and the wearable body-contact operating envelope (up to 45\,$^\circ$C); 0\,$^\circ$C is included to characterize the low-temperature boundary, though it lies outside the intended deployment range, as discussed in Section~\ref{sec:results:pvt}.
VCO characterization is performed open-loop by sweeping $C$ across the four targeted frequency values.

\textit{Device Sizing}: Table~\ref{tab:sizing} lists component values for the 300\,kHz variant. For lower-frequency variants, only the VCO load capacitor $C$ and the loop-filter passive values change; all transistor widths, lengths, and VCO resistor values are retained, confirming the reusability of the topology.

\begin{table}[t]
\centering
\caption{Device sizing for the 300\,kHz PLL variant.}
\label{tab:sizing}
\resizebox{\linewidth}{!}{%
\setlength{\arrayrulewidth}{0.4pt}
\begin{tblr}{
  colspec = {Q[60] Q[60] Q[80]},
  vline{2,3} = {2,3,4,5,6,7,8,9,10,11,12,13,14,15,16}{0.4pt},
  rowsep = 1.1pt,
  hline{1,2,14} = {1.2pt},
  hline{3,6,8,12} = {-}{0.4pt},
  hline{4,5,7,9,10,11,13} = {2,3}{0.4pt},
}
\textbf{Block} & \textbf{Element} & \textbf{Value / Size} \\
\SetCell[r=1]{l}\textbf{PFD}
  & \SetCell[c=2]{c} All cells from FlexIC PDK \\
\SetCell[r=3]{l}\textbf{CP}
  & R0:R1   & 50\,k$\Omega$ \\
  & T0:T1   & W=2\,\textmu m, L=600\,nm \\
  & \SetCell[c=2]{c} INV from FlexIC PDK \\
\SetCell[r=6]{l}\textbf{VCO}
  & R0      & 50\,k$\Omega$ \\
  & R1:R7   & 172.5\,k$\Omega$ \\
  & R8      & 120\,k$\Omega$ \\
  & R9      & 5.6\,M$\Omega$ \\
  & T0:T2   & W=2\,\textmu m, L=600\,nm \\
  & $C$     & 0.27\,pF \\
\SetCell[r=2]{l}\textbf{LPF}
  & $R_1$   & 50\,k$\Omega$ \\
  & $C$     & 5\,pF \\
\end{tblr}
}
\end{table}

\subsection{VCO Characterisation}
\label{sec:results:vco}

\begin{figure}[t]
    \centering
    \includegraphics[width=\linewidth]{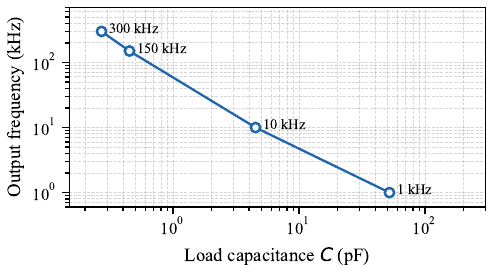}
    \caption{Open-loop VCO output frequency versus load capacitance
             $C$ (log-log scale). Each annotated marker corresponds
             to one co-design variant. The near-linear relationship
             confirms $C$-dominated delay across the full
             1\,kHz--300\,kHz range.}
    \label{fig:VCO} 
\end{figure}

Fig.~\ref{fig:VCO} shows the open-loop VCO output frequency as a function of load capacitance $C$.
A monotonic relationship spanning three decades---1\,kHz to 300\,kHz---is achieved by varying $C$ from 0.27\,pF to 52\,pF, a range of approximately 200$\times$, while keeping all other component values fixed.
On a log-log scale, the curve is near-linear, confirming that the delay-capacitance relationship of the ring oscillator stages is dominated by the capacitance $C$---a prerequisite for predictable co-design retargeting.
The four operating points each lie precisely on the curve at the required target frequency. 

\subsection{Steady-State Performance}
\label{sec:results:ss}

Table~\ref{tab:perf} summarizes steady-state performance at the 300\,kHz operating point (27\,$^\circ$C, $V_\mathrm{DD}=3$\,V, TT corner), the most demanding variant in terms of VCO speed and loop dynamics.
Total power of 115.4\,\textmu W is dominated by the VCO (110.6\,\textmu W), which carries the only unavoidable static bias current in the topology; the PFD and CP+LPF together contribute just 4.8\,\textmu W, reflecting the near-zero quiescent path in the charge pump and the purely passive low-pass filter.
This power distribution is a deliberate design outcome: by assigning all static bias to the VCO and eliminating quiescent currents in all other blocks, the topology minimises both total power and power-area tradeoffs.
The RMS period jitter of 2.24\,ns (0.07\,\% of the output period) confirms that the closed-loop feedback effectively bounds short-term timing uncertainty.
The long-term frequency accuracy of 1000\,ppm reflects a residual static offset (300.1\,kHz vs.\ the ideal 300\,kHz target) attributable to charge-pump current mismatch; crucially, this is a bounded, constant error in sharp contrast to the unbounded drift of a free-running RO, which accumulates indefinitely over time.

\begin{table}[t]
  \centering
  \caption{Simulated PLL performance (300\,kHz variant,
           $V_\mathrm{DD}=3$\,V, TT, 27\,$^\circ$C).}
  \label{tab:perf} 
  \renewcommand{\arraystretch}{1.25}
  \begin{tabular}{lcc}
    \toprule
    \textbf{Parameter} & \textbf{Value} & \textbf{Unit} \\
    \midrule
    Supply voltage $V_\mathrm{DD}$  & 3       & V          \\
    Output frequency                & 300     & kHz        \\
    Total power                     & 115.4   & \textmu W  \\
    \quad VCO                       & 110.6   & \textmu W  \\
    \quad PFD                       & 0.2     & \textmu W  \\
    \quad CP + LPF                   & 4.6     & \textmu W  \\
    Area      & 0.0115 & mm$^2$     \\
    RMS period jitter               & 2.24    & ns         \\
    Long-term freq.\ accuracy       & 1000    & ppm        \\
    \bottomrule
  \end{tabular} 
\end{table}

\subsection{PVT Robustness}
\label{sec:results:pvt}

\begin{figure}[t]
    \centering
    \includegraphics[width=\linewidth]{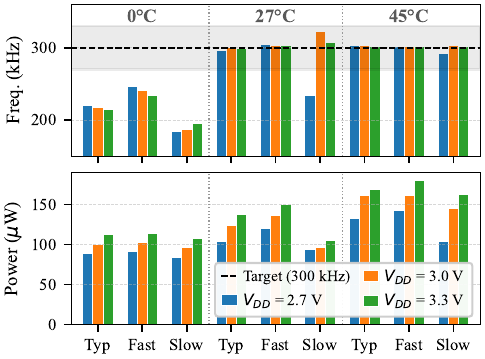}
    \caption{Locked output frequency (top) and total power (bottom) across 27 PVT combinations: three temperatures (0\,$^\circ$C, 27\,$^\circ$C, 45\,$^\circ$C) $\times$ three process corners (Typ/Fast/Slow) $\times$ three supply voltages (2.7, 3.0, 3.3\,V). Dashed line and shaded band indicate the 300\,kHz target $\pm$10\,\%. At 0\,$^\circ$C IGZO mobility is insufficient to reach 300\,kHz; the PLL acquires lock reliably at 27\,$^\circ$C and 45\,$^\circ$C.}
    \label{fig:PVT} 
\end{figure}

Fig.~\ref{fig:PVT} reports locked frequency and power across 27 PVT combinations. 
A key material property of amorphous IGZO is a positive temperature coefficient of electron mobility: carrier mobility---and hence VCO oscillation frequency---increases with temperature.
This creates a strong temperature dependence of the lock point that must be understood when characterising robustness.

At 0\,$^\circ$C the VCO runs 17--39\,\% below the 300\,kHz target across all process corners and supply voltages, regardless of VDD, placing the operating point entirely outside the loop's capture range. 
The PLL does not lock at 0\,$^\circ$C for the 300\,kHz variant. This is a physically motivated limitation rooted in the IGZO temperature coefficient of mobility, not in the PLL topology itself.
The intended deployment range of this design is 27--45\,$^\circ$C, consistent with wearable body-contact applications where skin-surface temperatures typically remain within this range~\cite{turn0search0}.
For applications requiring sub-ambient operation, the VCO could be rebiased or a larger $C$ selected to shift the lock point to lower frequencies, at the cost of a different target frequency.

At 27\,$^\circ$C---nominal room temperature---the PLL locks in 8 of 9 corner combinations: all three supply voltages at the typical and fast process corners lock within $\pm$7.5\,\% of the target (the worst case being 322.4\,kHz at 3.0\,V fast, +7.5\,\%), well within the $\pm$10\,\% tolerance band.
The single failing corner (2.7\,V, slow process, 233.9\,kHz, $-$22\,\%) represents an extreme combination of minimum supply and maximum process spread; in a realistic deployment at rated supply (3.0\,V) the PLL locks reliably across all process corners.
The supply voltage range of 2.7--3.3\,V corresponds to a $\pm$10\,\% variation around the nominal 3\,V, which covers the typical supply tolerance of flexible battery-powered wearables.

At 45\,$^\circ$C—the upper bound of typical wearable and device thermal stress tests—all 9 corners lock within $\pm$2.4\,\% of the 300\,kHz target, with a maximum deviation of $+$1.3\,\% at typical and fast corners.
Testing at this elevated temperature confirms robustness across realistic thermal excursions that flexible wearables may experience during operation or localized heating; device exterior surfaces in contact with human skin are often designed to remain below 43–48\,\textdegree C for safety and comfort~\cite{turn0search0}.

Power scales monotonically with both supply voltage and temperature, ranging from 84\,\textmu W at the lowest corner (0\,$^\circ$C, 2.7\,V, slow) to 181\,\textmu W at the highest (45\,$^\circ$C, 3.3\,V, fast), a variation of approximately 2.1$\times$ across the full PVT space.
In the primary operating region (27--45\,$^\circ$C, 3.0\,V nominal) the power range is 96--175\,\textmu W, confirming predictable power-performance scaling.

\subsection{Evaluation across target frequencies}
\label{sec:results:scaling}

Table~\ref{tab:scaling} reports PLL performance across the four application-frequency variants.
Retargeting is achieved solely by changing $C$ from 0.27\,pF (300\,kHz) to 52\,pF ($\approx$1\,kHz), a factor of $\approx$200$\times$ in capacitance; all transistor sizes, resistor values, and the PFD and CP sub-blocks are unchanged.
Power grows modestly from 115.4\,\textmu W to 152.9\,\textmu W (+32\,\%) with increasing capacitance, as the larger $C$ raises the static bias current needed to sustain oscillation at the lower frequency.
This moderate power increase---over a 300$\times$ range of target frequency---confirms that the static bias topology efficiently amortizes VCO energy over the output period regardless of frequency.
RMS jitter scales naturally with the output period: from 0.002\,\textmu s at 300\,kHz (0.07\,\% of period) to 12.7\,\textmu s at 1\,kHz (1.3\,\% of period).
Both bounds are within the timing margins reported for all four host systems, confirming that the PLL's residual jitter does not constrain system performance across any target application.

\begin{table}[t]
  \centering
  \caption{PLL performance across the four co-design variants
           ($V_\mathrm{DD}=3$\,V, TT, 27\,$^\circ$C).} 
  \label{tab:scaling}
  \renewcommand{\arraystretch}{1.25}
  \begin{tabular}{lcc}
    \toprule
    \textbf{Target} &
    \textbf{Power (\textmu W)} &
    \textbf{Jitter$_\mathrm{rms}$ (\textmu s)} \\
    \midrule
    \cite{FlexRISCV:MICRO25} 300\,kHz
      & 115.4 & 0.002 \\
    \cite{FlexCoProc:DATE26} 150\,kHz
      & 131.2 & 0.018 \\
    \cite{FlexClassifier:ICCAD25} 10\,kHz
      & 150.9 & 0.198 \\
    \cite{Alkhalil:BioCAS:2022:FlexibleSAR} $\approx$1\,kHz
      & 152.9 & 12.7  \\
    \bottomrule
  \end{tabular} 
\end{table}

\begin{figure}[t]
  \centering
  \includegraphics[width=\linewidth]{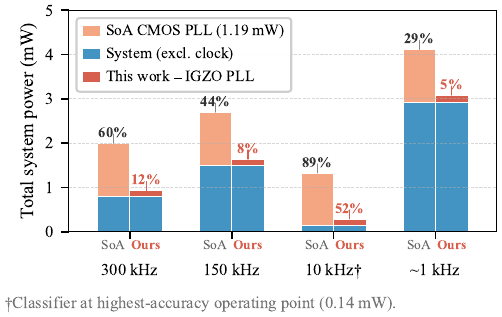}
  \caption{Absolute system power when each target is clocked by a SoA IoT CMOS PLL~\cite{Liu:TCSI2017:IoTPLL} (left bar) versus the proposed IGZO PLL (right bar). The system power (blue) is identical in both bars; only the clock contribution differs. Percentages above each bar indicate the clock's share of total power.}
  \label{fig:motivation} 
\end{figure}

Although no flexible PLL exists in the literature, a designer requiring a stabilized clock in an IGZO AMS platform today would have no choice but to integrate an external CMOS solution; accordingly, we include a SoA IoT CMOS PLL~\cite{Liu:TCSI2017:IoTPLL} as a baseline to quantify the power cost of this only available alternative in IGZO platforms.
Fig.~\ref{fig:motivation} compares the system power breakdown when each target is clocked by the proposed IGZO PLL versus the CMOS reference.
With the CMOS PLL, the clock alone accounts for 29--89\,\% of total system power; the proposed IGZO PLL reduces this to 5--12\,\% for three targets: 12\,\% for~\cite{FlexRISCV:MICRO25}, 8.0\,\% for~\cite{FlexCoProc:DATE26}, and 5.0\,\% for~\cite{Alkhalil:BioCAS:2022:FlexibleSAR}.
Across all targets, the CMOS clock overhead is up to 5.8$\times$ higher than the proposed IGZO PLL.
For~\cite{FlexClassifier:ICCAD25}, we report the
highest-accuracy operating point (0.2\,mm$^2$, 0.14\,mW), where the proposed PLL contributes 52\,\% of system power versus 89\,\% for the CMOS alternative.

The PLL area varies across the four variants (Table~\ref{tab:scaling}),
driven primarily by the size of the load capacitor $C$.
When normalized to the total system area (Table~\ref{tab:targets}),
the PLL contribution remains modest: approximately 1.1--2.0\,\%
for~\cite{FlexRISCV:MICRO25}, 0.47\,\%
for~\cite{FlexCoProc:DATE26}, 0.05--6.3\,\%
for~\cite{FlexClassifier:ICCAD25}, and 0.25\,\%
for~\cite{Alkhalil:BioCAS:2022:FlexibleSAR}.
These results confirm that closed-loop clock stabilization introduces only a minor overhead, making it a tractable and low-cost building block for complete flexible systems.

\subsection{Comparison with Prior Clocking Approaches}
\label{sec:results:comparison}

Table~\ref{tab:comparison} compares the proposed design against reported oscillator-based clocking solutions in FE.
Prior works employ free-running ROs or open-loop VCOs operating at 5--25\,V and consuming 0.4--50\,mW---one to three orders of magnitude above the 0.115--0.153\,mW of this design.
None reports jitter or frequency accuracy, since these metrics are not meaningful for open-loop oscillators with inherently unbounded drift.
Three distinctions stand out.
First, this is the only FE clocking solution operating at 3\,V, directly matching the supply of all target IGZO AMS systems and eliminating any level-shifting overhead.
Second, it is the only solution providing bounded frequency accuracy (1000\,ppm), replacing open-loop drift with reference-anchored stability---a property that is simply absent, not merely worse, in all prior FE works.
Third, our solution presents the smallest footprint, with a reduction of more than 1500$\times$ w.r.t. VCO alternatives, and more than 390$\times$ w.r.t. RO solutions.

\begin{table}[t]
  \centering
  \caption{Comparison with all published clocking solutions in FE.
           $^\dagger$VCO, $^\ddagger$RO, $^\star$PLL.}
  \label{tab:comparison}
  \renewcommand{\arraystretch}{1.0}
  \begin{tabular}{lccc}
    \toprule
    \textbf{Work} &
    \textbf{Freq (kHz)} &
    \textbf{Pwr (mW)} &
    \textbf{Area (mm$^2$)} \\
    \midrule
    \cite{VCO_igzo_tejaswini}$^\dagger$ & 0.33--0.56  & 1.3          & N/R    \\
    \cite{VCO_Bongjun}$^\dagger$        & 0.85--2.05  & $\approx$0.4 & 17.5    \\
    \cite{high-speed_yuanfeng}$^\ddagger$& 1820--6510 & $\approx$43  & 0.13    \\
    \cite{A-igzo_di}$^\ddagger$         & 810         & $\approx$50  & 0.208   \\
    \cite{fastiIC_arun}$^\ddagger$      & 406--2100   & 13--43       & 1.6--4.5  \\
    \midrule
    \textbf{This work}$^\star$
      & \textbf{1--300}
      & \textbf{0.115--0.153}
      & \textbf{0.0115--0.0233}\\
    \bottomrule
  \end{tabular}
\end{table}
\section{Conclusion}\label{sec:conclusion}

This paper presented the first PLL designed for n-type-only IGZO thin-film transistor technology, targeting clock stabilization in flexible AMS wearable systems.
The proposed design consumes 115.4\,\textmu W at 3\,V, achieves an RMS jitter of 2.24\,ns and a bounded frequency accuracy of 1000\,ppm, and occupies 0.0115--0.0233\,mm$^{2}$ depending on the integration target---up to $435\times$ reduction in power than prior flexible clocking solutions, and the first to provide reference anchored frequency stability.
A single topology, retargeted solely through VCO capacitor scaling, covers four published IGZO systems spanning 1\,kHz to 300\,kHz, validating frequency co-design as a practical and reusable framework.
PVT simulation confirms lock across the full wearable operating envelope. 
Since existing flexible clocking solutions are inadequate and optimized clock integration can jeopardize system feasibility, a key finding is that clock generation is not a peripheral overhead in IGZO platforms: integrating an unoptimized clock source can consume up to 90\,\% of total system power.
The proposed IGZO PLL mitigates this by reducing clock power to at most 52\,\% across four published target platforms---up to 5.8$\times$ lower than a SoA CMOS alternative.
Depending on the integration context, the PLL area ranges from 0.0115\,mm$^{2}$ to 0.0233\,mm$^{2}$, corresponding to approximately 1.1--2.0\,\%, 0.47\,\%, 0.05--6.3\,\%, and 0.25\,\% of total system area for the four target platforms.
These results establish on-chip PLL-based clocking as a viable, low-cost building block for the next generation of fully integrated, energy-efficient flexible wearable systems.

\section*{Acknowledgment}

This work has been supported by the European Research Council (ERC) (Grant No. 101052764).

\bibliographystyle{IEEEtran}
\bibliography{references}

\end{document}